\documentclass[journal=jpcafh,manuscript=article]{achemso}
\setkeys{acs}{chaptertitle = true}

\usepackage[T1]{fontenc}
\usepackage{amsmath,amssymb}
\usepackage{siunitx}
\usepackage[version=4]{mhchem}
\usepackage{graphicx}
\usepackage{booktabs}
\usepackage{makecell}
\usepackage{rotating}
\usepackage{threeparttable}
\usepackage{textcomp}
\usepackage{gensymb}
\usepackage{cleveref}
\usepackage{listings}
\usepackage{caption}
\usepackage{tabularx}
\usepackage{multirow}
\usepackage{chemformula}

\graphicspath{ {./figures/} }

\author{Esther Ritov}
\affiliation{Wolfson Department of Chemical Engineering, Technion -- Israel Institute of Technology, Haifa 3200003, Israel}
\author{Alon Grinberg Dana}
\affiliation{Wolfson Department of Chemical Engineering, Technion -- Israel Institute of Technology, Haifa 3200003, Israel}
\alsoaffiliation{Grand Technion Energy Program (GTEP), Technion -- Israel Institute of Technology, Haifa 3200003, Israel}
\alsoaffiliation{Resnick Sustainability Center of Catalysis, Technion -- Israel Institute of Technology, Haifa 3200003, Israel}
\email{alon@technion.ac.il}
\phone{+972-73-378-2117}

\title{
  Predictive Chemical Kinetic Modeling of Pt-Catalyzed Dry Methane Reforming
}

\begin{document}

\begin{abstract}
Dry reforming of methane (DRM) over platinum catalysts offers a promising route for \ce{CO2} utilization and syngas (\ce{H2}/\ce{CO}) production, a versatile feedstock for synthetic fuels. This study employs automated chemical kinetic model generation to present a detailed microkinetic mechanism to investigate Pt-catalyzed DRM between 700-1100 K, identifying key reaction pathways and kinetic limitations through sensitivity analysis.
Model predictions of \ce{CH4} and \ce{CO2} conversion and syngas production closely match fixed‐bed experimental data across the entire temperature range and varied feed ratios.
Our predictive reaction network reveals that \ce{OCX} serves as a critical bottleneck intermediate for cooperative \ce{CH4} and \ce{CO2} activation. While \ce{CO} desorption (\ce{OCX <=> CO + X}) was identified as the most influential step with a strong negative sensitivity coefficient toward methane concentration, \ce{OCX} regeneration (\ce{CO2X + CX <=> 2OCX}) inhibits methane conversion by maintaining surface saturation.
Methane activation follows sequential C–H scissions (\ce{CH4X -> CH3X -> CH2X -> CHX -> CX}), while \ce{CO2} activation proceeds predominantly via a hydrogen‐mediated carboxyl route (\ce{CO2X + HX -> COOHX -> COX + OHX}).
Three operational regimes were identified: low-temperature desorption-limited kinetics (\ce{OCX} saturation, 700-850~K), transitional pathway activation (site liberation, 850-950~K), and high-temperature distributed control with carbon management challenges (950-1300~K).
These insights provide design principles for platinum-based DRM catalysts, emphasizing the importance of promoting beneficial reactions while suppressing kinetic traps and implementing temperature-specific optimization strategies to balance conversion efficiency with catalyst stability. To our knowledge, this is the first study to apply fully automated, rule-based microkinetic model generation and experimental validation to Pt-catalyzed DRM, integrating kinetic sensitivity, flux analysis, and surface speciation for catalyst design. 

\end{abstract}
\clearpage

\section{1. Introduction}
\label{introduction}

The escalating global climate crisis has placed greenhouse gas (GHG) emissions at the forefront of environmental concerns, with emissions over the last decade reaching unprecedented levels in human history \cite{IPCC2022}.
One of the promising and extensively studied approaches to reduce net GHG emissions is dry reforming.
This catalytic process converts carbon dioxide (\ce{CO2}) and methane (\ce{CH4}), into synthesis gas (syngas), a mixture of hydrogen (\ce{H2}) and carbon monoxide (\ce{CO}) \cite{Li2023}.
Syngas serves as a crucial feedstock for the production of carbon-based synthetic fuels such as sustainable aviation fuel (SAF), synthetic diesel, or oxygenated hydrocarbons (e.g., methanol, dimethyl ether, and oxo-synthesis products) \cite{Wang2024}.

The underlying dry reforming of methane (DRM) mechanism involves four principal steps: 
(i) dissociative adsorption of methane, occurring preferentially at step‐edge sites with \ce{CH_x} fragments adopting site‐specific binding geometries\citep{wei2004mechanism}; 
(ii) dissociative adsorption of \ce{CO2} that proceeds rapidly and is strongly influenced by the surface structure and defect sites\citep{wittich2020catalytic}; 
(iii) hydrogen migration from the metal to the support, leading to hydroxyl formation, particularly below 800~\textdegree{}C, that is associated with a fast water–gas shift reaction\citep{Prins2012};  
(iv) oxidation and desorption of intermediates such as \ce{CH_x}, resulting in \ce{CO} formation either directly or via carbonate‐mediated reduction pathways\citep{Zhu2024}.

Despite extensive investigations, the precise surface reaction pathways remain a subject of debate, largely influenced by the catalyst structure and operating conditions.
Meanwhile, practical implementation of DRM continues to encounter significant challenges.\cite{Gao2023}
It is a highly endothermic process that requires relatively high temperatures (800-1000\textdegree{}C) to overcome the stability of the reactant, i.e., the high dissociation energies of \ce{CH4} and \ce{CO2}.\cite{Chen2023}
While catalysts play a critical role in reducing activation energies, maintaining efficiency under these extreme conditions remains challenging \cite{Zheng2023}.
Several undesirable side reactions occur during the DRM process, further complicating its feasibility \cite{PourAli2023}.
Two significant side reactions that contribute to catalyst deactivation are \ce{CH4} cracking and the Boudouard reaction\cite{Kumar2023}.
These reactions lead to the accumulation of carbon on the catalyst, significantly reducing reaction rates and accelerating deactivation\cite{Liu2023}.

The Boudouard reaction~\eqref{eq:Boudouard} predominates at lower temperatures (250-350~\textdegree C), while methane cracking~\eqref{eq:Cracking} becomes significant at higher temperatures (600-800~\textdegree C)~\cite{nist2025}. 

\begin{equation}
\label{eq:Boudouard}
2\,\ce{CO_{(g)}} \rightleftharpoons \ce{C_{(s)}} + \ce{CO2_{(g)}} \qquad \Delta H^\circ_{298\,\text{K}} = -172.4~\text{kJ/mol}
\end{equation}

\begin{equation}
\label{eq:Cracking}
\ce{CH4_{(g)}} \rightarrow \ce{C_{(s)}} + 2\,\ce{H2_{(g)}} \qquad \Delta H^\circ_{298\,\text{K}} = +74.9~\text{kJ/mol}
\end{equation}

The Boudouard reaction is a non-elementary process that proceeds via a three-step surface mechanism involving intermediate species~\cite{diblasi2009}. First, \ce{CO2} dissociatively adsorbs on a carbon-free active site, forming a surface-bound carbon–oxygen complex and releasing \ce{CO}. In the second step, the surface oxygen species reacts with adsorbed carbon to produce another \ce{CO} molecule, regenerating the active site. The third step describes the reverse process, in which \ce{CO} re-adsorbs and reacts with surface oxygen to reform \ce{CO2}.
Likewise, methane cracking is also non-elementary and proceeds through dissociative adsorption on metal surfaces~\cite{en15072573}. The mechanism involves a sequence of surface-mediated dehydrogenation steps, where methane is incrementally stripped of hydrogen atoms, ultimately leading to the deposition of elemental carbon.

These reactions increase the likelihood of coke formation, ultimately reducing the longevity and efficiency of the catalyst.\cite{Abdullah2022, Zhang2022}
Therefore, catalyst development is essential to improve DRM performance.
The primary challenge is designing economically viable catalysts while demonstrating high activity, superior selectivity, and strong resistance to deactivation.\cite{Aramouni2018}

Catalysts used for DRM can be classified into earth-abundant transition metals and noble metals.\cite{Ghani2019}
Among the transition metals, nickel (\ce{Ni}) is widely studied due to its cost-effectiveness and high activity.
However, Ni-based catalysts are prone to coking because they readily promote the carbon-forming side reactions mentioned above.\cite{Yu2023}
In contrast, noble metals, particularly platinum (\ce{Pt}), exhibit superior stability under oxidative conditions, significantly reducing carbon formation \cite{Gokon2022}.
Although noble metals offer significantly better deactivation resistance than transition metals, their high cost remains a drawback.

A deeper understanding of the elementary steps that govern the activation and conversion of a \ce{CO2} and \ce{CH4} mixture is essential to optimize the catalyst design and reaction conditions \cite{Gao2023}.
Developing predictive chemical kinetic models\cite{GreenPredictive} that are based on fundamental reaction steps and justifiable parameters can provide valuable insights into material performance and process efficiency.\cite{Chen2023, Li2023}
Such advancements will establish a pivotal framework for designing catalysts that minimize deactivation caused by sintering and coking at the high temperatures (>800\textdegree{}C) required for substantial methane and carbon dioxide conversion during DRM\cite{Pakhare2014, Awad2024}, while enhancing our understanding of reaction pathways across temperature ranges to optimize catalytic activity \cite{Kathiraser2015, Niu2021}.

This study employs predictive chemical kinetic modeling to investigate Pt-catalyzed DRM between 500-850\textdegree{}C.
The present work aims to identify key reaction pathways and catalyst poisoning species, providing insights into temperature-dependent carbon formation and informing strategies for improved catalyst design.

\section{2. Methods}
\label{methods}

The chemical kinetic model was generated using the Reaction Mechanism Generator (RMG) software version 3.2.0~\cite{RMG3, RMGdb, RMGCat2017}. 
RMG automatically explores possible intermediate species and elementary reactions for a given reacting mixture and thermodynamic conditions and utilizes a flux-based algorithm that considers the net production rates of species.
The procedures implemented by RMG to iteratively expand the kinetic mechanism are extensively discussed elsewhere~\cite{RMG3, RMGdb, RMGDocs, Gao2016RMG, GreenPredictive}.
Briefly, RMG automatically explores possible intermediate species and elementary reactions for a given reacting mixture using template-based rules and pre-compiled libraries.
Given the initial temperature, pressure, and reactant composition, together with surface binding energies and site density, the software estimates all unknown reaction rate coefficients and species thermodynamic properties using rate rules\cite{RMGdb} and the group additivity method\cite{Banson1969GAV}.
The user sets the model expansion tolerance, which serves as the 'importance' criterion for species inclusion in the model by the flux-based algorithm.

Previously, RMG successfully generated a kinetic reaction mechanism for the heterogeneous catalysis of methane dry reforming on nickel~\cite{GoldsmithWest2017}.
The present mechanism was constructed using RMG’s thermodynamic libraries for Pt(111) species~\cite{blondal2019computer,kreitz2022detailed}, 
the gas-phase \ce{H2}/\ce{O2} system~\cite{BurkeH2O2}, 
CCSD(T)-F12 and QCISD(T) based computations~\cite{buesser2022thermoDFT, goldsmith2012thermoQCI, harper2022thermo}, 
and RMG’s kinetic libraries for
high-temperature catalytic partial oxidation of methane over platinum~\cite{quiceno2006modeling, vlachos2007catalytic}. 
Simulations were performed at temperatures of 600-1300 K (~330-1030\textdegree{}C), a pressure of \SI{1}{bar}, and inlet \ce{CH4}/\ce{CO2} mole ratio values between 0.2-1.5.
A termination rate ratio of 0.01 was used and the expansion tolerance was set at 0.05~\cite{RMGDocs}.

After generating the reaction mechanism with RMG, the DRM process was simulated using a catalytic reactor model in Cantera~\cite{cantera}. A plug-flow reactor (PFR) configuration was employed, represented as a series of perfectly mixed, constant-volume, zero-dimensional reactors with negligible pressure drop along the reactor length.

Two different reactor configurations were employed to validate the model against experimental datasets. In the first configuration, matching the Gustafson1991 study~\cite{Gustafson1991}, the catalyst bed was assigned a surface-area-to-volume ratio of \SI{1000}{\per\metre}, corresponding to a monolith reactor whose central section was coated with catalytically active Pt. Operating conditions were established at a pressure of \SI{1}{bar}, with a reactor diameter of \SI{25.4}{\milli\metre} (1~inch), a reactor length of \SI{0.1}{\metre}, and an inlet flow rate of \SI{102}{\cubic\centi\metre\per\minute} at standard temperature and pressure (STP), yielding a calculated residence time of approximately 5.9~seconds. The model assumes an idealized Pt(111) surface, represented by a single site type (X), with a standard site density of $2.483 \times 10^{-9}~\mathrm{mol\,cm^{-2}}$.

The second configuration was designed to replicate the experimental conditions reported by Niu et al.~\cite{niu_drm_2021}, using the same RMG libraries with the addition of argon to the species database. Minor modifications were applied to the RMG mechanism generation parameters, including a surface-area-to-volume ratio of \SI{1000}{\per\metre} and a termination time of 2~seconds. Subsequently, the Cantera reactor configuration was adapted to align precisely with the reported experimental setup. The reactor geometry was scaled to an internal diameter of \SI{10}{\milli\metre} and a length of \SI{12.7}{\milli\metre}, while the catalyst surface-area-to-volume ratio was increased to \SI{1500}{\per\metre}. The total volumetric flow rate was set to \SI{200}{\milli\litre\per\minute}, with an inlet composition of 15\% \ce{CH4}, 15\% \ce{CO2}, and 70\% Ar by volume, maintaining a 1:1 stoichiometric ratio between \ce{CH4} and \ce{CO2}. The gas hourly space velocity (GHSV) was calculated as \SI{180000}{\milli\litre_{CH4}\per\hour\per\gram_{cat}}, based on the methane flow rate and a catalyst mass of \SI{10}{\milli\gram}. Temperature-dependent simulations were performed across the experimental temperature range of 450-800~\si{\celsius} (723-1073~K), with the reactor discretized into 7001 segments to ensure sufficient spatial resolution.

No adjustments were made to the kinetic or thermodynamic model parameters. The model presented and discussed here is purely predictive.\cite{GreenPredictive}

\section{3. Results and Discussion}

\subsection{3.1. Benchmarking the model}

\begin{figure}[htbp]
  \centering
  \includegraphics[width=0.9\textwidth]{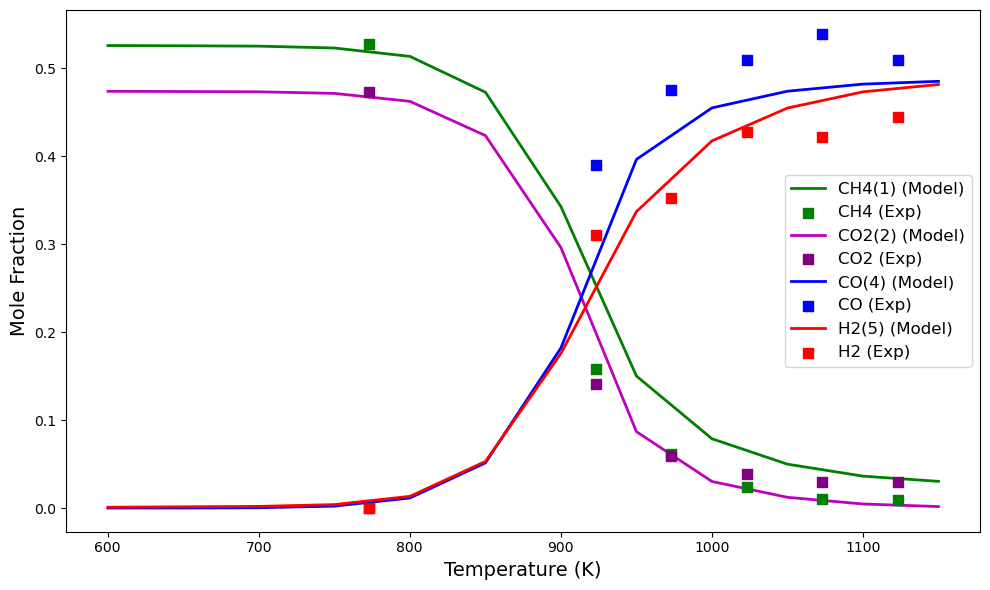}
  \caption{Comparison between experimental data (symbols)~\cite{Gustafson1991} and model predictions (solid lines, this work) for Pt-catalyzed DRM with a \ce{CO2}/\ce{CH4} feed ratio of 0.9. The upper panel presents mole fractions of \ce{CH4}, \ce{CO2}, \ce{CO}, and \ce{H2} as a function of temperature. The lower panel shows conversion percentages of \ce{CH4} and \ce{CO2} over the 700-1123\,K range (500-850\textdegree{} C), with a residence time of 5.9~s.}
  \label{fig:Updated_Model}
\end{figure}

\begin{figure}[htbp]
  \centering
  \includegraphics[width=0.9\textwidth]{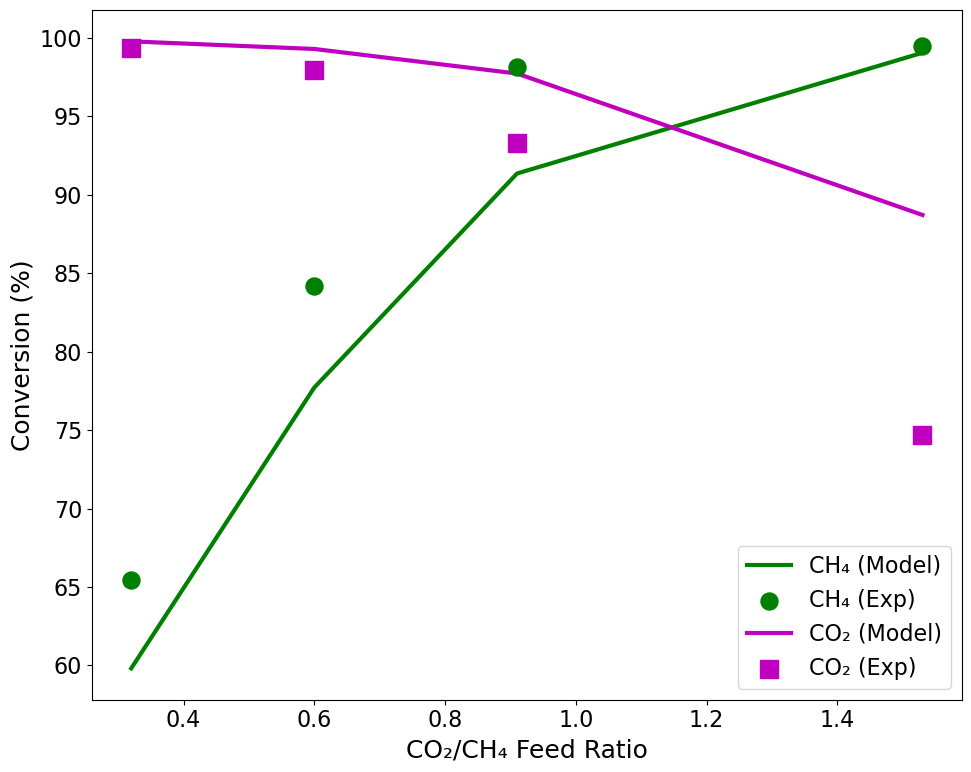}
  \caption{Conversions of \ce{CH4} and \ce{CO2} as a function of the \ce{CO2}/\ce{CH4} feed ratio at 1123\,K, with a residence time of 5.9~s.}
  \label{fig:24.04.25_Conversion}
\end{figure}

The influence of temperature and feed composition on the DRM process was experimentally studied by Gustafson and Walden~\cite{Gustafson1991} using a fixed-bed reactor. Our model predictions, benchmarked against their data, are presented in Figures~\ref{fig:Updated_Model} and \ref{fig:24.04.25_Conversion}, highlighting the effects of temperature and reactant feed ratios, respectively.

Figure~\ref{fig:Updated_Model} illustrates the temperature dependence of species mole fractions over the 700-1123\,K range. The model successfully reproduces the experimentally observed trends for \ce{CH4}, \ce{CO2}, \ce{CO}, and \ce{H2}. It predicts the onset of the DRM reaction at around 800\,K, and aligns reasonably well with the experimental observation. Throughout the entire temperature range, the model maintains an approximate equimolar ratio of \ce{CO} to \ce{H2}, with a slightly higher concentration of \ce{CO}, as expected from the 0.9 \ce{CO2}/\ce{CH4} feed ratio. While a minor over-prediction of unconverted \ce{CH4} is observed at higher temperatures relative to \ce{CO2}, the overall conversion trends are captured satisfactorily well by the model.

Figure~\ref{fig:24.04.25_Conversion} compares the modeled and experimental reactant conversions at 1123\,K as a function of the \ce{CO2}/\ce{CH4} feed ratio. The model captures the observed profiles reasonably well, including the decline in methane conversion and the corresponding increase in syngas production with increasing feed ratio. The model slightly over-predicts the conversion of \ce{CO2} at the high \ce{CO2}/\ce{CH4} feed ratio. Each reactant achieves its maximum conversion at a feed ratio corresponding to a reduced concentration of that reactant in the mixture. Across all ratios, the conversion of the limiting reactant consistently exceeds that of the excess reactant.

\begin{figure}[htbp]
  \centering
  \includegraphics[width=0.9\textwidth]{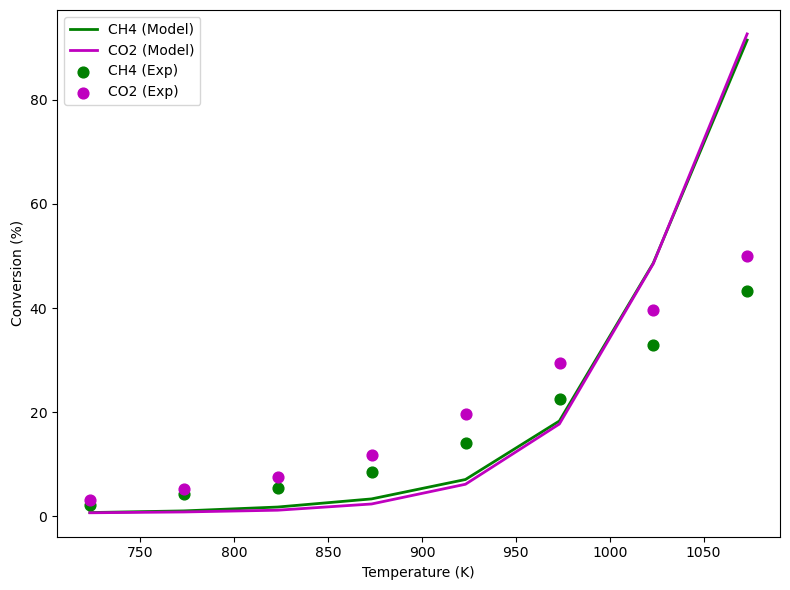}
\caption{Comparison between experimental data (symbols)~\cite{niu_drm_2021} and model predictions (solid lines, this work) for Pt-catalyzed dry reforming of methane (DRM) under a \ce{CH4}/\ce{CO2}/Ar feed ratio of 0.15/0.15/0.70 at 1\,bar. The plot shows conversion percentages of \ce{CH4} and \ce{CO2} as a function of temperature in the range of 700-1100\,K, with a residence time of 0.3\,s.}
\label{fig:Niu_Model}
\end{figure}

Figure~\ref{fig:Niu_Model} presents validation against an independent dataset under different conditions, including a shorter residence time (0.3~s), equimolar feed ratio, and dilute operation with argon. The model maintains a consistent predictive trend; however, at high temperatures, it overpredicts the conversion of both \ce{CH4} and \ce{CO2} compared to the experimental data. This validation, combined with the Gustafson and Walden benchmarking, demonstrates the mechanism's robustness across diverse operating conditions and confirms its reliability for industrial applications and catalyst design.

\subsection{3.2. Flux analysis}

\begin{figure}[h!]
    \centering
    \includegraphics[width=0.8\textwidth]{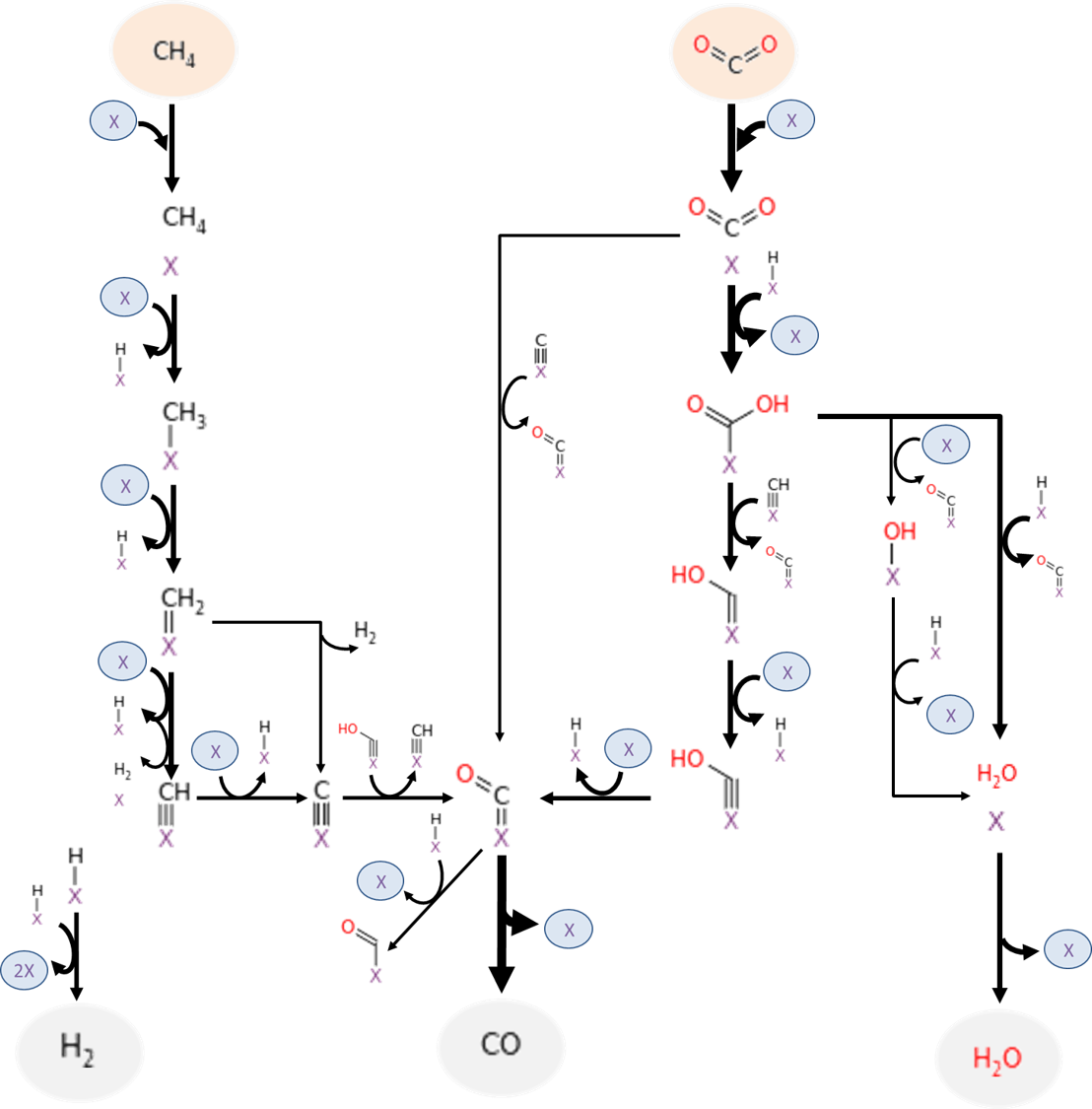}
    \caption{Detailed reaction network for the DRM process over a platinum catalyst.\ce{CO2}/\ce{CH4} feed ratio=0.9, $\tau = 5.9$~s }
    \label{fig:Flux}
\end{figure}

Figure~\ref{fig:Flux} illustrates the detailed reaction network for the DRM process over a platinum catalyst at 1000~K. The diagram reveals an interconnected pathway architecture governed by surface adsorption and molecular activation, highlighting the major pathways responsible for hydrogen evolution and intermediate coupling.

The reaction network predominantly follows \ce{C1}-based pathways, in line with the mechanistic framework established for platinum-catalyzed methane activation~\cite{aghalayam2003,zhang2021,wittich2020}.
Methane activation proceeds via a stepwise \ce{C-H} bond scission sequence on platinum surface sites, captured in the dehydrogenation progression:
\begin{equation*}
\ce{CH4X -> CH3X -> CH2X -> CHX -> CX}.
\end{equation*}

Among these, the dissociation of \ce{CHX} to \ce{CX} and \ce{HX} (\ce{CHX <=> CX + HX}) is frequently identified as a kinetically significant barrier, based on both kinetic modeling and Density Functional Theory (DFT) studies~\cite{niu2016dry}. This fourth \ce{C-H} cleavage is often considered rate-limiting in methane activation on transition metals due to the high activation energy required to rupture the final \ce{C-H} bond on the surface.

Its central position in the flux diagram-bridging the \ce{CH_x} intermediates with surface-bound carbon formation-highlights its role as a mechanistic junction between upstream dehydrogenation and downstream oxidation pathways. Each dehydrogenation step concurrently generates adsorbed hydrogen atoms (\ce{HX}), which desorb via recombination:
\begin{equation*}
\ce{2HX <=> H2 + 2X}.
\end{equation*}

This step constitutes the primary pathway for molecular hydrogen formation, visible in the left region of the reaction network, and is essential for maintaining active site availability and sustaining the reforming cycle.

On the \ce{CO2} side, activation predominantly follows a hydrogen-assisted pathway:
\ce{CO2X + HX <=> COOHX}. This route exhibits a significantly lower activation barrier (72.0~kJ/mol) compared to direct \ce{CO2} dissociation (174.5~kJ/mol)~\cite{niu2016dry}, underscoring the thermodynamic and kinetic favorability of the \ce{COOHX} intermediate. The importance of this pathway is also reflected in the reaction flux diagram, where the carboxyl route dominates over alternatives.

Li~et~al.~\cite{li2024ptceo2} observed characteristic bands between 1000-2000~cm$^{-1}$ corresponding to O-C-O vibrational modes on Pt/CeO$_2$ catalysts, which they attributed to adsorbed carboxyl (\ce{COOH}), formate (\ce{HCOO}), and carbonate species. Similarly, Mierczynski~et~al.~\cite{mierczynski2023ftir} demonstrated through FTIR spectroscopy that CO2 on Ni/CeO2 catalysts can be hydrogenated to form COOH intermediates, which subsequently undergo further hydrogenation to bicarbonates, supporting the hydrogen-assisted CO2 activation pathway.

Following its formation, the carboxyl intermediate undergoes C-O bond cleavage via
\ce{COOHX <=> COX + OHX}, facilitating the generation of \ce{CO}. In contrast, the competing formate pathway (\ce{HCOOX}) is more endothermic and associated with higher energy barriers, making it kinetically less favorable and thus less significant in the overall mechanism.

The carboxyl intermediate plays a central role, facilitating C-O bond scission to yield \ce{CO} (\ce{COOHX <=> COX + OHX}). In contrast, the formate pathway (\ce{HCOOX}, not shown) involves higher barriers and is more endothermic, making it kinetically less favorable. Since \ce{HX} species are a major product of hydrocarbon dehydrogenation, the hydrogen-mediated \ce{CO2} activation pathway serves as a key coupling junction between the \ce{CH4} and \ce{CO2} sub-networks. This synergy is further enhanced by hydrogen spillover, where surface hydrogen generated from \ce{CH4} dehydrogenation migrates across the surface to promote \ce{CO2} activation. The phenomenon of hydrogen spillover on platinum-containing catalysts is well-established~\cite{khoobiar_h_spillover_1964, im_h_spillover_pt_zeolite, karim_h_spillover_support_effects}, and its role in facilitating DRM-relevant processes has been directly demonstrated~\cite{li_h_spillover_drm}.

A key mechanistic feature revealed in the network is the direct reaction between surface carbon species and adsorbed carbon dioxide, \ce{CX + CO2X <=> 2COX}, which serves as another crucial coupling junction between the degradation pathways of the two primary reactants. This Langmuir-Hinshelwood mechanism~\cite{Liu2017} proceeds via the reaction of two surface-bound species, \ce{CX} and \ce{CO2X}, and enables the conversion of inert carbon deposits into mobile \ce{COX} species, playing a dual role in both \ce{CO} formation and catalyst reactivation.

The central region of the flux diagram captures the sequential oxidation pathways of adsorbed methylidyne species (\ce{CHX}). Our analysis indicates that the oxidation of \ce{CHX} predominantly proceeds via \ce{COOH}-mediated oxygen transfer, as exemplified by the elementary step \ce{CHX + COOHX -> CH2OX + OCX}.
This pathway, involving the reaction between an adsorbed \ce{CHX} and an adsorbed carboxyl intermediate (\ce{COOHX}), leads to the formation of a partially oxidized carbon species (\ce{CH2OX}) and adsorbed carbon monoxide (\ce{OCX}). While specific experimental verification of this exact elementary step is complex to isolate directly in DRM on platinum, numerous DFT studies on carbon oxidation and oxygen transfer mechanisms on platinum group metals provide strong theoretical support for the high reactivity of surface carboxyl species as oxygen donors, and for their preference over hydroxyl-mediated pathways for oxidizing carbonaceous species~\cite{scaranto2016, gao_hcooh_pt111, lin_hcooh_pt111, wang_chx_oxidation_pt13}. This finding highlights the crucial role of oxygen-containing surface species, specifically \ce{COOHX}, in oxidizing carbonaceous intermediates within the reaction network.

The preferential reactivity of carboxyl intermediates in carbon oxidation processes on platinum surfaces is well-supported by comprehensive DFT studies~\cite{scaranto2016, gao_hcooh_pt111, lin_hcooh_pt111}, which demonstrate that \ce{COOH} dehydrogenation proceeds readily on Pt(111) and that carbon oxidation products, such as \ce{CO} and \ce{CO2}, are primarily formed through \ce{COOH} intermediate pathways in related reactions such as formic acid decomposition. Furthermore, the significant role of \ce{COOH} intermediates in the mechanism of the water-gas shift reaction on Pt catalysts has been established through first-principles calculations and microkinetic modeling~\cite{grabow2008}. Microkinetic modeling efforts for methane reforming and related reactions on platinum-group metals have also emphasized the importance of various intermediates in their proposed reaction mechanisms~\cite{khandekar2017}.

The interconnected nature of these pathways underscores how platinum catalysts achieve high DRM performance through coordinated activation of both reactants. Collectively, the reaction network reveals critical trade-offs between activity and selectivity. Computational validation supports the flux preferences and intermediate distributions shown. The centrality of \ce{COOH} intermediates, the role of hydrogen spillover and hydrogen-assisted mechanisms, and the kinetic accessibility of key elementary steps emerge as defining factors that govern DRM performance over Pt catalysts.

\subsection{3.3. Kinetic Sensitivity Analysis}

\begin{figure}[!htb]
\centering
\includegraphics[width=0.9\textwidth]{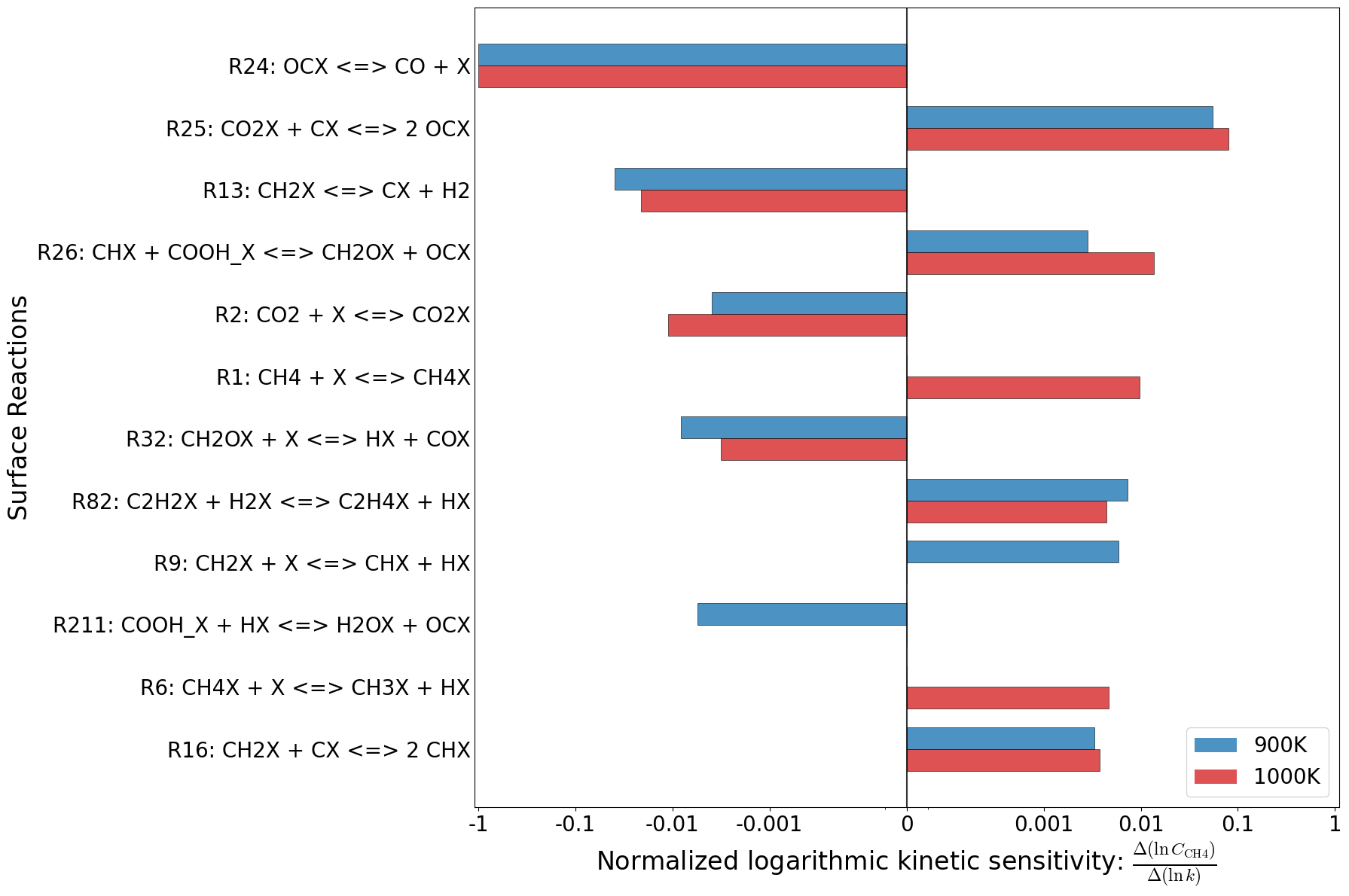}
\caption{Normalized kinetic sensitivity coefficients, shown on a logarithmic scale, for the concentration of \ce{CH4} in DRM over a platinum catalyst at 900~K and 1000~K, highlighting the temperature-dependent shift in rate-determining steps. Negative sensitivity coefficients indicate reactions that promote methane consumption when the respective rate coefficient slightly increases, and vice versa.}
\label{fig:Kinetic_SA}
\end{figure}

The kinetic sensitivity analysis of DRM over a platinum catalyst, performed at 900~K and 1000~K with a \ce{CO2}/\ce{CH4} feed ratio of 0.9 and a residence time of 5.9 seconds, illustrates temperature-dependent reaction dynamics and pinpoints the critical pathways governing methane conversion.
These two temperatures represent about 70\% and 20\% conversion of \ce{CH4}, respectively (Figure~\ref{fig:Updated_Model}).
The normalized sensitivity coefficients, shown in Figure~\ref{fig:Kinetic_SA}, quantify the effect of perturbations in individual rate coefficients on methane concentration, providing insights into the controlling reaction pathways and how their influence shifts with temperature.
By examining normalized sensitivity coefficients, we gain quantitative insights into how key individual elementary surface reactions influence the overall process, focusing here (Figure~\ref{fig:Kinetic_SA}) on the concentration of methane as the observable.

\begin{figure}[htbp]
    \centering
    \includegraphics[width=0.8\textwidth]{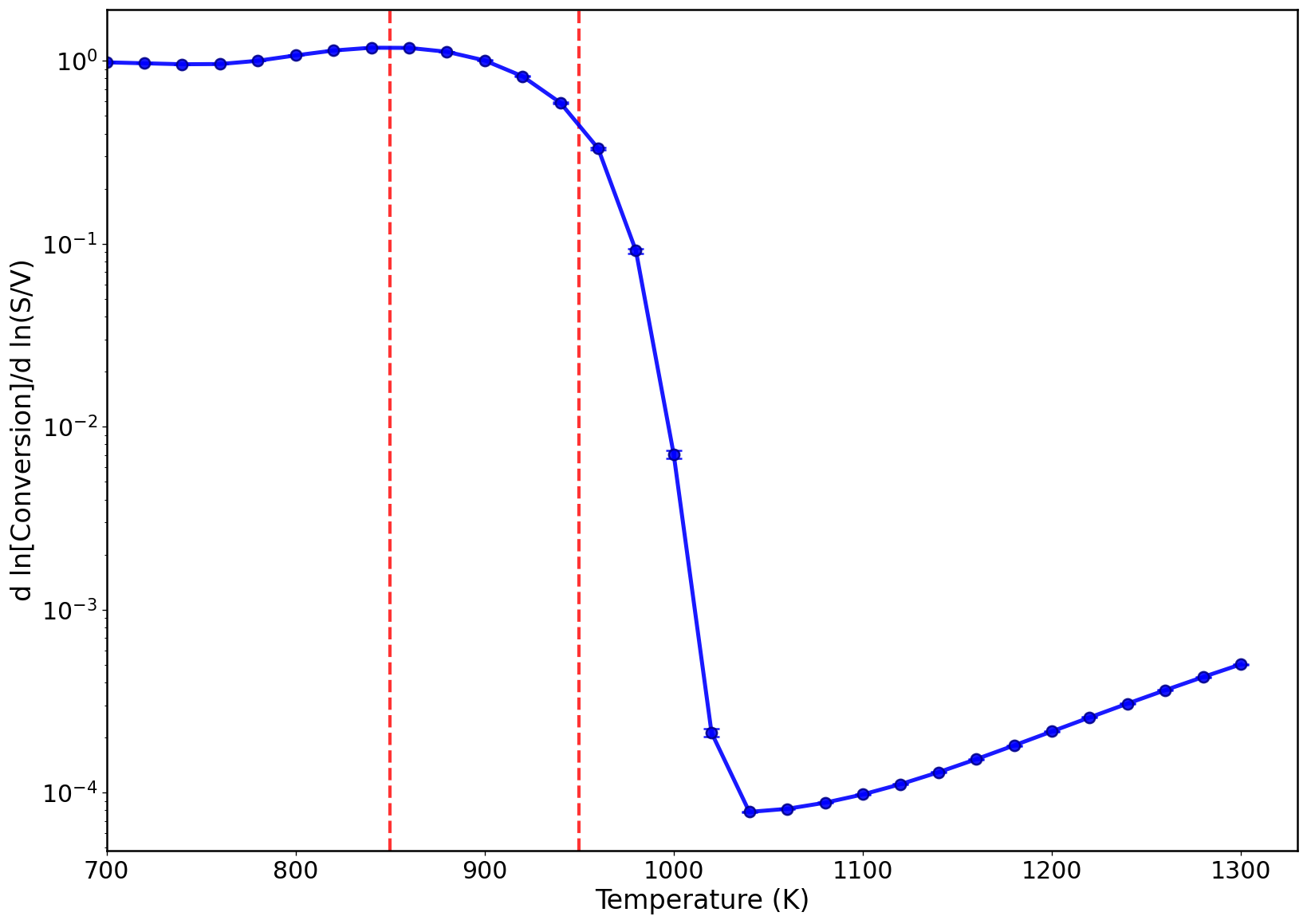}
    \caption{Temperature dependence of methane conversion sensitivity to variations in the surface area to volume ratio (S/V). The normalized sensitivity coefficient $\frac{d \ln[\text{Conversion}]}{d \ln(S/V)}$ was computed by perturbing the S/V ratio by $+$5\% around a baseline value of 1000 m$^{-1}$. The y-axis is shown on a logarithmic scale. Vertical dashed lines indicate key transition boundaries near 850\,K and 950\,K.}
    \label{fig:sv_sensitivity}
\end{figure}

Figure~\ref{fig:sv_sensitivity} presents the normalized sensitivity coefficients for methane conversion with respect to the catalyst's surface area to volume ratio (S/V), plotted on a logarithmic scale as a function of temperature. This metric quantifies the direct impact of catalyst surface availability on the overall reaction efficiency. The resulting curve reveals a strongly temperature-dependent regime characterized by three distinct behaviors.

Below approximately 950\,K, the sensitivity remains high and nearly constant, maintaining values close to 1. This indicates that the conversion is critically limited by the available catalytic surface area, consistent with the trends seen in total site coverage (Figure~\ref{fig:clean_total_occupied_sites_linear_y_vs_temp}). In this regime, the catalyst surface is predominantly saturated with adsorbed intermediates such as \ce{CHX}, \ce{OCX}, and \ce{HX} (Figure~\ref{fig:individual_coverage_log}). These strongly-bound species limit the number of free active sites, and thus increasing the surface area directly alleviates the kinetic bottleneck, resulting in enhanced conversion.

Between approximately 950\,K and 1000\,K, the sensitivity exhibits a sharp and abrupt decline, spanning over four orders of magnitude (from $\sim 10^{0}$ to $\sim 10^{-4}$). This dramatic transition reflects a major reorganization in surface kinetics, driven by an increase in temperature that facilitates desorption and site liberation. As a result, the system's dependence on the macroscopic S/V ratio diminishes rapidly. This indicates a shift in the rate-limiting step from site-limited adsorption/reaction to regimes potentially dominated by gas-phase mass transfer or intrinsic surface kinetics, where turnover frequencies are no longer constrained by surface availability.

Above approximately 1000\,K, the sensitivity gradually increases again, although it remains very low ($\sim 10^{-4}$ to $\sim 10^{-3}$). This subtle rise suggests a weak re-emergence of S/V dependency. At these elevated temperatures, rapid desorption kinetics allow for fast regeneration of surface sites. Thus, even small increases in surface area can still improve conversion marginally by increasing the density of these rapidly cycling sites. This effect could also be linked to the activation of new, more demanding high-temperature pathways requiring specific surface interactions.

Further support for these interpretations comes from the kinetic sensitivity analysis (Figure~\ref{fig:Kinetic_SA}), which highlights the changing role of key elementary steps. Notably, \ce{OCX <=> CO + X} (R24) and recombination reactions (e.g., R211) become increasingly relevant with temperature, influencing \ce{CH4} conversion and contributing to the observed regime transitions.

Reaction R24, \ce{OCX <=> CO_{(g)} + X}, exhibits the highest absolute normalized sensitivity coefficient value, displaying a strong negative sensitivity at both temperatures, indicating that increasing the \ce{CO} desorption rate coefficient has the strongest effect on promoting \ce{CH4} consumption.
The \ce{OCX} species serves as a critical rate determining step in the reaction network (Figure~\ref{fig:Flux}), serving as a key intermediate that connects multiple competing pathways. The extreme negative sensitivity coefficient of R24 (an order of magnitude larger than all other sensitivity coefficients in absolute terms) reveals that CO desorption from surface-bound \ce{OCX} represents the primary kinetic bottleneck in the catalytic cycle. When this desorption step is accelerated, it rapidly clears \ce{OCX} intermediates from the surface, freeing active sites and driving the overall reaction network forward.

This depletion effect is balanced by the compensating reaction R25, \ce{CO2X + CX <=> 2OCX}, which shows strong positive sensitivity at both temperatures. R25 regenerates \ce{OCX} by consuming both \ce{CO2X} and surface carbon (\ce{CX}). Its positive sensitivity indicates that accelerating this regeneration pathway actually inhibits the overall methane conversion. The sensitivity analysis thus reveals a delicate optimization balance: sufficient \ce{CO} desorption (R24) must occur to prevent kinetic bottlenecks, but excessive \ce{OCX} regeneration (R25) can destabilize the catalytic cycle.

The dehydrogenation step R13, \ce{CH2X <=> CX + H2} displays strong negative sensitivity toward \ce{CH4} concentration, indicating its critical role in promoting methane conversion at both examined temperatures. In the flux network (Figure~\ref{fig:Flux}), both \ce{CH2X} and \ce{CX} are major intermediates, and the negative sensitivity coefficient of R13 suggests that enhancing its rate accelerates the overall reaction flux by efficiently converting methane-derived carbon species to valuable hydrogen product.
This reaction serves a dual role: it produces \ce{CX}, which can serve as a carbon reservoir, and \ce{H2}, a desired product. The network interpretation reveals that R13 acts as a carbon flux regulator that maintains the balance between carbon utilization and hydrogen production. Its consistently negative sensitivity coefficient at both temperatures identifies R13 as a pivotal control point for both performance optimization and carbon management in the catalytic system.

Reactions R2, \ce{CO2 + X <=> CO2X}, and R32, \ce{CH2OX + X <=> HX + OCX}, exhibit negative sensitivity coefficients at both temperatures, indicating that they consistently promote methane conversion across the operating range. R2 represents \ce{CO2} adsorption that facilitates the overall conversion process, contrary to intuitive expectations of site competition. The negative sensitivity of R2 at both temperatures suggests that \ce{CO2} adsorption creates favorable surface conditions or provides essential \ce{CO2X} intermediates that enhance the reforming efficiency. R32 involves the transformation of \ce{CH2OX} intermediates, generating surface hydrogen (\ce{HX}) and carbonyl species (\ce{OCX}). Its negative sensitivity coefficient at both temperatures indicates that accelerating this transformation promotes methane conversion by facilitating intermediate processing and maintaining optimal surface speciation.

Four reactions exhibit temperature-dependent sensitivity patterns that reveal the evolving kinetic landscape. Reactions R82, \ce{C2H2X + H2X <=> C2H4X + HX}, R9, \ce{CH2X + X <=> CHX + HX}, and R211, \ce{COOH_X + HX <=> H2OX + OCX}, show negative sensitivity primarily at 900~K, indicating that they promote methane conversion under lower-temperature conditions. These reactions represent alternative pathways that become kinetically relevant when the main reaction network operates at moderate activity levels. 

R9 provides an alternative \ce{CH2X} transformation route that competes with the dominant R13 pathway, while R82 involves C–C coupling reactions that may contribute to carbon utilization. R211 represents a hydrogen-assisted intermediate coupling that generates water and regenerates \ce{OCX}. At 1000~K, these alternative pathways become kinetically overshadowed by the dominant routes, demonstrating temperature-dependent pathway activation. These four reactions display clear differences in their sensitivity coefficients between the two temperatures (900K vs 1000K), either by appearing at only one temperature or showing significantly different magnitudes at the two temperatures.

The emergence of R1, \ce{CH4 + X <=> CH4X}, exclusively at 1000~K with positive sensitivity reflects a fundamental shift in methane activation behavior and represents the fourth temperature-dependent reaction. At 900~K, the low sensitivity of R1 indicates that conventional methane activation routes, \ce{CH4 -> CH3X -> CH2X -> CHX}, are sufficient due to favorable kinetics and adequate surface site availability. However, at 1000~K, R1 emerges with positive sensitivity, suggesting that this major pathway begins to compete with or interfere with more efficient routes. The positive sensitivity indicates that while R1 provides an additional methane activation mechanism, its enhancement inhibits overall conversion, possibly due to formation of less reactive \ce{CH4X} species that act as kinetic dead-ends or occupy surface sites unproductively.

Additional temperature-dependent reactions emerge at 1000~K with distinct sensitivity patterns. Reaction R6, \ce{CH4X + X <=> CH3X + HX}, has a positive sensitivity coefficient exclusively at 1000~K, indicating that while this reaction attempts to reactivate surface carbon species by converting them to \ce{CH3X}, its acceleration actually inhibits overall methane conversion. This suggests that the \ce{CH4X} species generated by R1 create a kinetic trap, where reactivation through R6 diverts surface sites and intermediates from more efficient pathways. R16: \ce{CH2X + CX <=> 2CHX} appears at both temperatures but shows enhanced (more negative) sensitivity at 1000~K, indicating it promotes methane conversion by efficiently recycling both methane-derived \ce{CH2X} and accumulated \ce{CX} into reactive \ce{CHX} intermediates. This reaction effectively doubles the availability of active carbon species, underscoring the importance of carbon recycling mechanisms at elevated temperatures where carbon management becomes critical for sustained catalytic performance.

\subsection{3.4. Implications for Catalyst Design and Side Reactions}

\begin{figure}[!htb]
\centering
\includegraphics[width=0.9\textwidth]{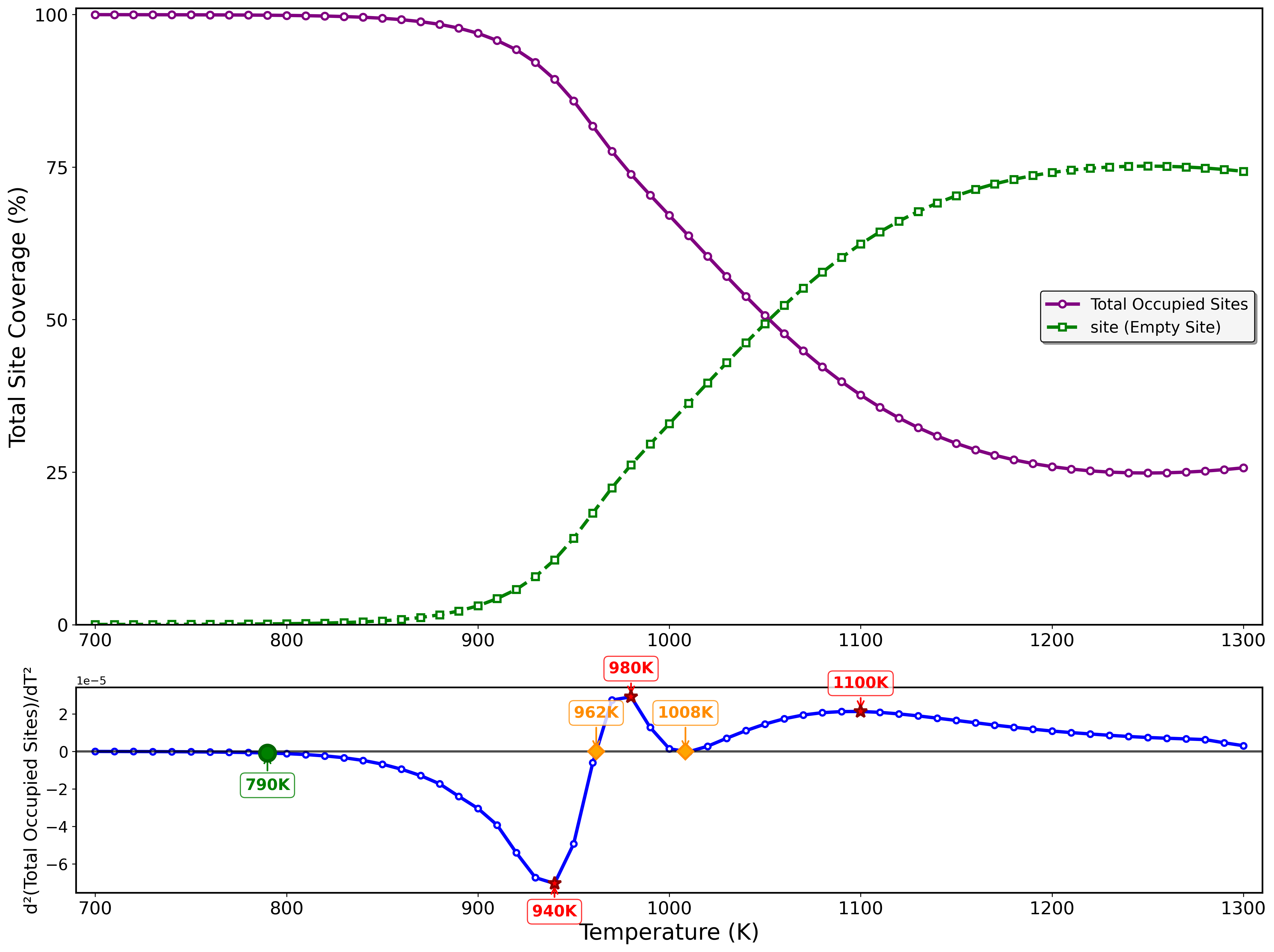}
\caption{Temperature-dependent surface site coverage and its second derivative. 
Top: Total occupied sites (purple) and empty sites (green) as a function of temperature reveal three operational regimes. 
Bottom: Second derivative of total site coverage (blue) identifies regime transitions: 
(i) surface saturation regime (790-980~K), with early onset detection at 790~K marking initial deviation from hydrogen-dominated saturation; 
(ii) transitional regime (980-1100~K), marked by zero crossings at 962~K and 1008~K (50\% site crossover); 
(iii) high-activity regime (1100-1300~K), where free sites dominate ($\sim$75\%) with near-zero curvature indicating thermodynamic stability.}
\label{fig:clean_total_occupied_sites_linear_y_vs_temp}
\end{figure}

\begin{figure}[!htb]
\centering
\includegraphics[width=0.9\textwidth]{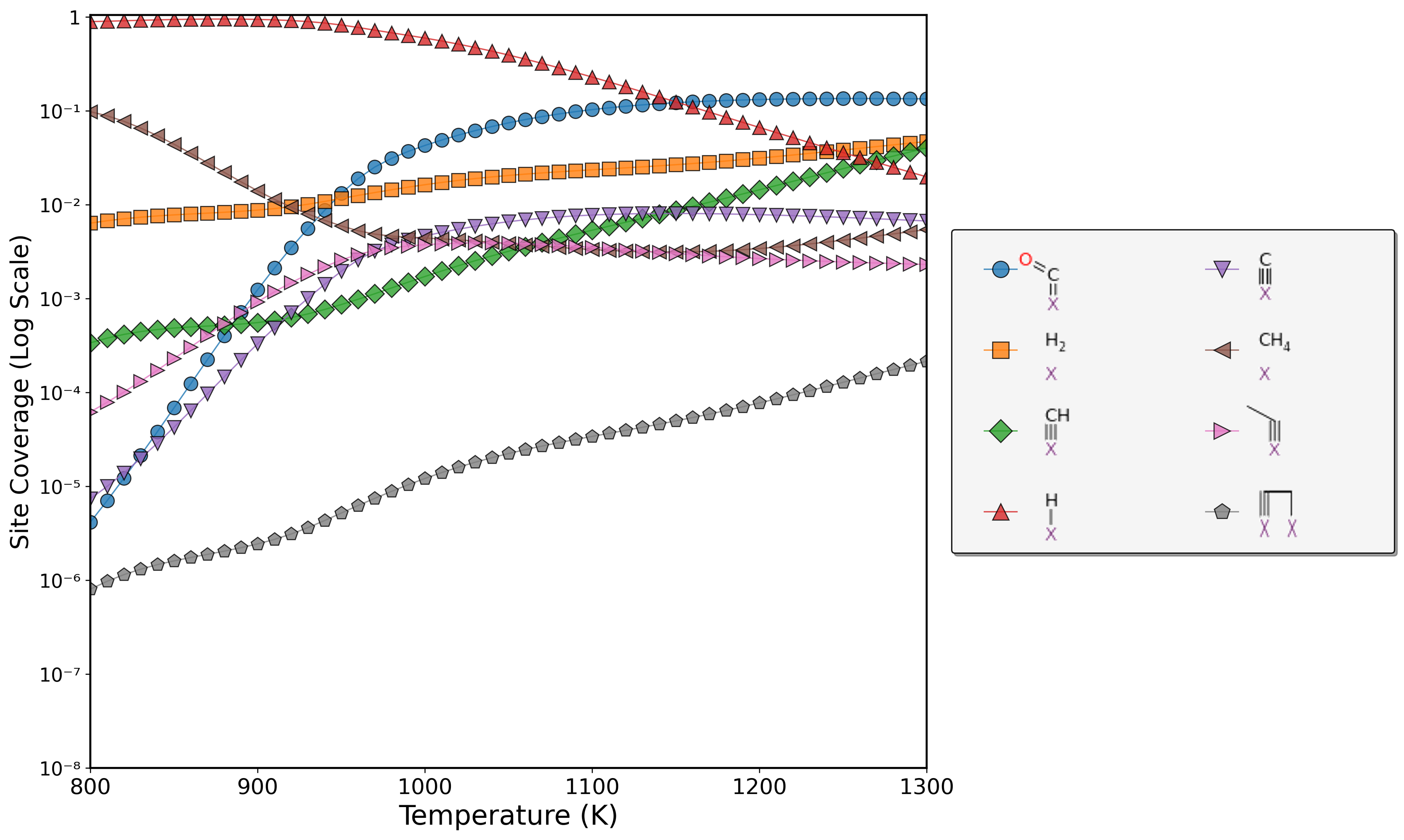}
\caption{Species-resolved surface coverage profiles across temperature. \ce{HX} (red triangles) dominates at low temperatures, \ce{OCX} (blue circles) peaks in mid-range, while \ce{CHX} (green diamonds) and \ce{CX} (purple right-pointing triangles) buildup dominates at high temperatures, reflecting transitions between different catalytic regimes.}
\label{fig:individual_coverage_log}
\end{figure}

Figure~\ref{fig:clean_total_occupied_sites_linear_y_vs_temp} traces the evolution of total surface coverage from 700 to 1300~K and reveals three distinct regimes traced by inflection points in the second derivative. At low temperatures (700-790~K), the surface is saturated ($\sim$100\% coverage), predominantly by adsorbed hydrogen species (\ce{HX}), as clearly demonstrated in Figure~\ref{fig:individual_coverage_log}. In this regime, free sites for methane activation are nearly absent due to hydrogen poisoning, and hydrogen desorption processes compete with other desorption steps for kinetic control, exhibiting a strong sensitivity in the reaction network.
The second derivative plot (bottom of Figure~\ref{fig:clean_total_occupied_sites_linear_y_vs_temp}) quantitatively delineates these regimes. The early onset detection at 790~K captures initial breakdown of hydrogen saturation. The most negative peak at 940~K in d$^2$(Total Occupied Sites)/d$T^2$ identifies the onset of rapid surface liberation, a critical temperature where hydrogen-dominated saturation collapses. The zero crossing at 962~K marks the deceleration of cooperative desorption, while the inflection at 1008~K coincides with the 50\% coverage crossover. The approach to near-zero curvature above 1100~K confirms thermodynamic stability.

Between 940~K and 1100~K, a transitional regime emerges where the surface undergoes dramatic speciation changes. The sharp desorption onset at 940~K initiates rapid \ce{HX} desorption, while \ce{OCX} coverage (blue circles) increases and reaches its peak around 1000-1100~K. Within this window, the total site coverage drops from $\sim$95\% to $\sim$25\%, activating methane-derived pathways: coverage of \ce{CHX} intermediates (green diamonds) emerges and increases substantially from $\sim 10^{-3}$ to $\sim 10^{-2}$, while \ce{H2X} species (orange squares) show sustained presence indicating active reforming. The critical crossover at 1008~K marks where occupied and empty sites reach equal coverage (50\% each). Kinetic control shifts from being dominated by hydrogen management to a more distributed network involving both C-H bond cleavage and \ce{CO2}-mediated oxidation.

Above 1100~K, the surface enters a high-activity regime characterized by sustained site availability ($\sim$75\% empty sites). The \ce{OCX} species reach their maximum and begin to plateau, while dramatic accumulation of surface carbon species occurs, most notably \ce{CX} (purple right-pointing triangles) which increases exponentially from $\sim 10^{-5}$ at 950~K to $\sim 10^{-3}$ at 1300~K, signaling potential for coke nucleation. The \ce{H2X} coverage achieves its highest levels ($\sim 10^{-1}$), indicating maximum hydrogen production capacity and strong reforming activity.

The emergence and peak of \ce{OCX} in the mid-temperature range confirms that \ce{CO2} dissociation becomes favorable as sites are liberated, with the sensitivity analysis at 1000~K revealing a dramatic redistribution of kinetic control. While \ce{CO} desorption (R24) remains negatively sensitive, its magnitude decreases as site availability improves, and methane dehydrogenation steps (R13: \ce{CH2X <=> CX + H2}, R6: \ce{CH4X <=> CH3X + HX}) develop positive sensitivities, confirming the activation of methane conversion pathways as \ce{HX} coverage declines and \ce{CHX} species emerge. The appearance of positive sensitivity for reforming-related reactions indicates that C-H bond cleavage and hydrogen management become kinetically relevant, consistent with the observed emergence of \ce{H2X} and \ce{CHX} species.

These trends support the multi-step Boudouard and methane cracking mechanisms described earlier. The \ce{OCX} dominance (blue circles) reflects initial \ce{CO2} dissociation steps, the emergence of \ce{CHX} (green diamonds) and \ce{HX} (red triangles) confirms sequential methane dehydrogenation, and \ce{CX} accumulation (purple down-pointing triangles) represents the final carbon products of both pathways. The progression from an \ce{HX}-dominated surface through the \ce{OCX} peak to \ce{CX} accumulation validates the mechanistic sequence in which hydrogen management transitions to carbon–oxygen chemistry and ultimately to carbon deposition pathways beginning around 950~K.

The exponential rise in \ce{CX} coverage from 950~K aligns with coke formation observed in TGA studies~\cite{das2020coke}, though the model captures only chemisorbed species. As shown in Figure~\ref{fig:individual_coverage_log}, \ce{CX} coverage (purple right-pointing triangles) increases dramatically from $\sim 10^{-5}$ to $\sim 10^{-3}$ between 950~K and 1300~K, representing a hundred-fold increase that signals the onset of coke precursor formation. Thus, the model provides valuable predictive insight into the early stages of coke nucleation, demonstrating the progression from hydrogen poisoning through active reforming to potential carbon accumulation, with carbon–carbon coupling reactions (R16: \ce{CH2X + CX <=> 2 CHX}) showing increased sensitivity in the high-activity regime.

To optimize catalyst performance across these regimes, different strategies are required based on the sensitivity-identified rate-controlling steps. In the saturation regime (790-980~K), promoting hydrogen desorption through support modification or metal-support interactions is essential to accelerate liberation from \ce{HX} dominance, while enhancing positive sensitivity reactions (R25, R1) can improve site utilization. In the transitional regime (980-1100~K), careful balance between \ce{OCX} formation and \ce{CHX} accumulation is needed, with precise temperature control around the critical 1008~K crossover. In the high-activity regime (above 1100~K), the exponential \ce{CX} growth necessitates carbon management strategies to prevent irreversible accumulation through carbon-recycling pathways and controlling carbon-carbon coupling reactions.

The species-resolved analysis reveals that surface poisoning mechanisms vary across regimes from hydrogen poisoning at low temperatures to potential carbon poisoning at high temperatures. This highlights the value of combining surface speciation analysis with kinetic sensitivity metrics to guide catalyst design for both high performance and long-term stability across the entire operational range.

\section{4. Conclusion}

In this work, we have developed and validated a detailed predictive microkinetic model for Pt‐catalyzed dry reforming of methane (DRM) over the 700-1300 K range using the Reaction Mechanism Generator (RMG) software.
The model captures reasonably well the temperature-dependent specie profiles measured across 700-1150 K, validated against observed species concentrations, and analyzed using flux and kinetic sensitivity data.
The suggested model also captures the temperature‐ and feed‐ratio–dependent methane and carbon‐dioxide conversions, as well as syngas product ratios, reproducing the onset of reforming near 800 K and the equimolar \ce{CO}/\ce{H2} trend at higher temperatures.

A comprehensive flux analysis at 1000 K reveals a tightly coupled network in which methane dehydrogenation proceeds through sequential C–H scissions (ultimately yielding \ce{CX}), while \ce{CO2} activation follows a hydrogen‐mediated carboxyl pathway.
Surface‐carbon oxidation is dominated by carboxyl transfer to \ce{CH_x} species and by direct reaction of \ce{CX} with \ce{CO2X} (Langmuir–Hinshelwood coupling), illustrating the pivotal role of \ce{COOH} intermediates and hydrogen spillover in maintaining catalytic turnover and in preventing coke formation.

The kinetic sensitivity analysis at 900 K ($\sim$70\% \ce{CH4} conversion) and 1000 K ($\sim$90\% conversion) identified the key rate‐controlling steps and their temperature dependence.
The \ce{CO} desorption reaction (\ce{OCX <=> CO + X}) is the most influential step, with extreme negative sensitivity coefficients, revealing desorption-limited kinetics under high surface coverage. Counterintuitively, \ce{OCX} regeneration (\ce{CO2X + CX <=> 2OCX}) shows positive sensitivity for \ce{CH4} concentration, indicating that excessive regeneration inhibits methane conversion by maintaining surface saturation and diverting intermediates from productive pathways.
The fourth C–H scission (\ce{CH2X <=> CX + H2}) is the most beneficial reaction for increased \ce{CH4} conversion at all studied conditions, making it a prime target for catalyst design.

Surface coverage analysis revealed three distinct regimes: (1) desorption-limited kinetics at low temperatures (700-850~K) with near-complete site saturation, (2) pathway diversification during site liberation (850-950~K), and (3) distributed kinetic control with carbon management challenges at high temperatures (950-1300~K).
In the \ce{COX}‐saturated regime, promoters or supports that lower the \ce{OCX} desorption barrier will be most effective.
In the transition regime, balancing \ce{CH_X} binding strength to sustain negative sensitivity in dehydrogenation is critical.
At high temperatures, catalysts must combine strong carbon‐recycling activity with suppression of \ce{C2}‐forming and deep‐hydrogenation routes to prevent deactivation.
These mechanistic insights establish design principles for platinum-based DRM catalysts that achieve both high conversion and operational stability through temperature-dependent kinetic control strategies.

The model predicts that CO* desorption initiates around 950–1000\,K, and identifies an optimal operating window between 1000–1100\,K where DRM activity is maximized while carbon accumulation is minimized, providing clear, experimentally testable targets for validation, and long-term catalyst stability tests.

Overall, our results demonstrate that a network‐level perspective, integrating flux analysis, sensitivity analysis, and surface speciation, can unambiguously identify the elementary steps that govern activity, selectivity, and stability in DRM.
These insights provide concrete guidelines for rational design of Pt‐based catalysts and reaction conditions that maximize syngas yields while mitigating coke formation under industrially relevant high‐temperature conditions.

\begin{suppinfo}

\begin{itemize}
  \item The RMG input file that was used to generate the model
  \item The RMG species dictionary
  \item The Chemkin format gas phase model
  \item The Chemkin format surface model
\end{itemize}

\end{suppinfo}

\begin{acknowledgement}

This research was supported in part by the Boeing-Technion SAF Innovation Center funded by the Boeing Company.
Financial support from the Stephen and Nancy Grand Technion Energy Program (GTEP) is gratefully acknowledged.
E.R. acknowledges funding through the Israel Scholarship Education Foundation (ISEF).

\end{acknowledgement}

\bibliography{bibfile}

\end{document}